\documentstyle[prl,aps,multicol,epsf]{revtex}
\def\dfrac#1#2{{\displaystyle{#1\over#2}}}

\begin{document}
\title{
Spin and orbital  excitation spectrum in the  Kugel-Khomskii model\\
%
}
\author{{G.~Khaliullin}\cite{Kh}}
\address{Max-Planck-Institut f\"ur Physik komplexer Systeme, D-01187 Dresden, Germany }
\author{{V.~Oudovenko}\cite{Udo}}
\address{Max-Planck-Institut f\"ur Festk\"orperforschung, 70569 Stuttgart, Germany}
\date{\today}

\maketitle

\begin{abstract}
We discuss spin and orbital ordering in the twofold orbital degenerate
superexchange model in three dimensions
 relevant to perovskite transition metal oxides.
We focus on the particular point on the classical
phase diagram where orbital degeneracy is lifted by
quantum effects exclusively.
Dispersion and damping of the spin and orbital excitations are
calculated at this point taking into account their mutual interaction.
Interaction corrections to the mean-field order parameters
are found to be small. We conclude that quasi-one-dimensional
N\'eel spin order accompanied by the uniform $d_{3z^2-r^2}$-type
orbital ordering is stable against quantum fluctuations.
\end{abstract}
\draft

\begin{multicols}{2}
\narrowtext

\newpage\


It is well known that the orbital (quasi)degeneracy
of $3d$-states in transition metal oxides plays an important role
in their magnetic and lattice properties.
An orbital ordering driven by exchange interactions and/or by
Jahn-Teller effect occurs at low temperature resulting in
a rich variety of magnetic structures (for a review see, e.g. \cite{KK}).
On the other hand a little is known, however,  on dynamical aspects
of the coupling between  spin and orbital degrees of freedom in these
systems:
a) What is the spectrum of low-energy orbital excitations,
b) How orbital excitations are coupled to the spin sector,
c) How this coupling  affects magnetic order parameter and
spin waves?
In the present paper we address these questions by considering
the superexchange model with twofold orbital degeneracy,
which corresponds to the $d^9$ Mott-Hubbard insulator on a cubic
lattice.

 To be specific, we consider the following Hamiltonian derived by
Kugel and Khomskii \cite{KK}, and studied recently by Feiner et al.
\cite{Oles}:

$$
H=\frac{t^2}{U}\sum_{\langle i j \rangle } \left [
4(\vec S_{i}\vec S_{j})
(\tau_{i}^{\alpha}-\frac{1}{2})
(\tau_{j}^{\alpha}-\frac{1}{2})+
\right.
$$
\begin{equation}
\left.
+(\tau_{i}^{\alpha}+\frac{1}{2})
 (\tau_{j}^{\alpha}+\frac{1}{2})-1
\right].
\end{equation}
In Eq.(1) we follow notations used in \cite{Oles}:
$t$ is the hopping between $e_g(3z^2-r^2) $ orbitals along the $c$-axis,
$\vec S_i$ is the spin $1/2$ operator. Operators $\tau_{i}^{\alpha}$ act in
the orbital subspace with basic vectors ${1\choose{0}}$, $ {0\choose{1}}$ corresponding to
the $e_g(x^2-y^2)\sim \mid x \rangle$ and $e_g(3z^2-r^2)\sim \mid z \rangle$
orbital states respectively.
The structure of $\tau_i^{\alpha}$ depends on the index $\alpha$ which
specifies the orientation of the bond $\langle ij \rangle$ relative to
the cubic axes $a,b,c$:

\begin{equation}
\tau_{i}^{a(b)}=\frac{1}{4}
(-\sigma_{i}^{z}\pm \sqrt{3}\sigma_{i}^{x}),\;\;\;\;\;
\tau_{i}^{c}=\frac{1}{2}\sigma_{i}^{z},
\end{equation}
where $\sigma^z$ and $\sigma^x$ are Pauli matrices.

It is rather easy to see that the classical N\'eel state
(i.e. $\langle \vec S_i \vec S_j \rangle= -1/4 $
in Eq. (1)) is infinitely degenerate: orbitals at each site
may be rotated independently. Feiner et al.~\cite{Oles} have
suggested that local orbital fluctuations associated with this degeneracy
strongly affect spin-sector when quantum fluctuations around the N\'eel
state are included, and drive the system into a disordered spin-liquid state
even in three dimensions~\cite{3}.
Our results presented below do not support this exciting scenario.
We have  investigated spin and orbital orderings,
and their excitations in the model defined by Eqs.~(1),(2).
Our main findings are $i)$ $\mid z \rangle$-type
orbital ordering favouring quasi-one-dimensional spin order is the most
promising candidate for the ground state.  $ii)$ Orbital excitations
have a gap generated by quantum effects.
This gap controls well the fluctuations around the mean-field solution.
$iii)$ Spin-orbit coupling does indeed act to decrease the staggered moment,
but this effect is not enough to destroy the long-range order
in a cubic lattice.

To begin with, we 
use the condition $\sum_{\langle i,j\rangle} \tau^{\alpha}_{i}=0$
following from Eq.~(2), and represent Eq.~(1) in a more transparent way:
\begin{equation}
H=-3+\sum_{\langle i,j \rangle } \hat J^{ij}_\alpha
( \vec S_i \vec S_j + \frac{1}{4}),
\end{equation}
\vspace*{-0.5cm}
\begin{equation}
\hat J^{ij}_{\alpha}=4 \tau_{i}^{\alpha} \tau_{j}^{\alpha} -
2(\tau_{i}^{\alpha}+  \tau_{j}^{\alpha} )+1.
\end{equation}

The first term in (3) represents the classical N\'eel energy
(in units of $t^2/U$) which we drop hereafter. From
the above Hamiltonian the key feature of the Kugel-Khomskii model
is evident: The exchange "constant" has in fact an internal operator
structure accounting for the orbital dynamics, and it's expectation
value strongly depends on the orientation of orbitals.
It follows from Eqs.~(3),(4) that the strength of the intersite
orbital coupling (hence the energy gain due to the orbital ordering)
is proportional to the deviation of spins from the  N\'eel state,
i.e. to the value of $\langle \vec S_i \vec S_j +\dfrac{1}{4} \rangle $.
This acts to reduce the effective dimensionality of the spin system:
Orbitals are arranged in  such a way which makes the exchange
coupling strongly nonuniform thus enhancing spin
fluctuations as much as possible.
In  low-dimensional models, a similar 
consideration suggest that the orbital ordering may lead to
the spin-liquid state~\cite{Pen}.
The $z$-type ordering of orbitals in the model~(1)
is suggested by this picture.
Indeed, the expectation value of exchange coupling (4) between
$z$ orbitals is $J_c=4$ along the $c$ axis, and it is only
small in the $(a b)$ plane: $J_{\perp}=1/4$.
Exchange energy is mainly accumulated in $c$ chains and
can be approximated as
$J_c \langle \vec S_i \vec S_j +\dfrac{1}{4} \rangle_c +
2J_\perp  \langle \vec S_i \vec S_j +\dfrac{1}{4} \rangle_\perp
\simeq -0.65 $ per site (using
$\langle \vec S_i \vec S_j\rangle_c =1/4 -\mbox{ln}2 $
for $1D$~\cite{Mattis} and assuming
$\langle \vec S_i \vec S_j\rangle_\perp \sim 0 $).
On the other hand $x$-type ordering results in the easy plane
magnetic structure ($J_{a,b}=9/4, J_c=0 $) with a much
smaller energy gain $\simeq -0.38$.

Our strategy is to study the Hamiltonian (3) within the
following scheme.
$i)$ We rewrite (3) in the form
$H=H_{sp} +H_{orb}+H_{int}$. Here the first two terms describe spin and
orbital sectors in the  mean-field level:

\begin{equation}
H_{sp}=\sum_{\langle i,j \rangle }
\langle  \hat J^{ij}_\alpha \rangle
( \vec S_i \vec S_j + \frac{1}{4}),
\end{equation}
\vspace*{-0.6cm}
\begin{equation}
H_{orb}=\sum_{\langle i,j \rangle }
 \langle  \vec S_i \vec S_j + \frac{1}{4}\rangle
\delta(\hat J^{ij}_\alpha),
\end{equation}
where $\delta A = A -  \langle  A \rangle $.
The crucial  importance  is   the stability of the
mean-field state against fluctuations generated by dynamical
coupling between spin and orbital excitations. This coupling
is represented by
\begin{equation}
H_{int}=\sum_{\langle i,j \rangle }
\delta (\hat J^{ij}_\alpha )
\delta (\vec S_i \vec S_j ).
\end{equation}
$ii)$ We assume the antiferromagnetic spin order and
uniform $z$- or $x$-type ordering of orbitals.
Then we employ spin (orbital)  wave representation for
$\vec S_i$ ($\vec \sigma_i$) operators.
$iii)$ We calculate spin-orbit interaction corrections  to the
excitation spectrum and to the order parameters.
Since latter quantities enter in  coupling constants
in Eqs.~(5),(6), all  steps have to be done in  the self-consistent way.

Consider $z$ orbital order which results in a highly  anisotropic
quasi-$1D$ magnetic structure.
We discuss first mean-field results,
which follow from Eqs.~(5),(6). 
Spin and orbital wave energies are given by
$\omega_{1k}=J_{1}\sqrt{1-\gamma_{1k}^{2}}$, and
$\omega_{2k}=J_{2}\sqrt{1+2\gamma_{2k}^{}}$ respectively.
Here
$J_{1}=(J_{c}+2J_{\perp})$, and
\begin{eqnarray}
J_{c}&=&\langle 1-\sigma_{i}^{z}-
\sigma_{j}^{z}+\sigma_{i}^{z}\sigma_{j}^{z}  \rangle_{c},\\\nonumber
J_{\perp}&=&\langle 1+\dfrac{1}{2}\sigma_{i}^{z}+\dfrac{1}{2}\sigma_{j}^{z}
+\dfrac{1}{4}\sigma_{i}^{z}\sigma_{j}^{z} +
\dfrac{3}{4}\sigma_{i}^{x}\sigma_{j}^{x}  \rangle_{\perp} .
\end{eqnarray}
The orbital stiffness is controlled by
$J_{2}=-8(\kappa_{c}-\dfrac{1}{4}\kappa_{\perp} )$,
\linebreak
 with
$\kappa_{\alpha}=\langle \vec S_{i}\vec S_{j}+\dfrac{1}{4} \rangle_{\alpha} $.
Momentum dependencies of $\omega_{nk}\;\;\;$
 (index $ n=1,2)$ are determined by the functions
\linebreak
$ \gamma_{1k}=(J_{c}\cos k_{z} +2 J_{\perp} \gamma_{k} )/J_{1} $,
$ \gamma_{2k}= -\dfrac{3}{2} \dfrac{\kappa}{4-\kappa}\gamma_{k}$,
where
$\kappa_{}= \kappa_{\perp}/\kappa_{c}$,
and
$ \gamma_{k}= \dfrac{1}{2} (\cos k_{x} + \cos  k_{y} ) $.
We calculate all expectation values within linear spin (orbital)
wave theory, with only one exception, the interchain spin
correlator which we approximate as 
\linebreak
$\langle \vec S_i \vec S_j \rangle_\perp =
\langle \vec S_i^z \rangle  \langle \vec S_j^z \rangle +
\langle S_i^{+} S_j^{-} \rangle $~\cite{Jperp}.

Self-consistent  mean-field calculations show that
the orbital pseudospin is almost saturated
(the mixture of $\mid x \rangle$ state is about one percent only).
Coupling between chains $J_\perp$ is weak (see Table 1)
but sufficient to produce quite large magnon dispersion in the
$(a b)$ plane (see thin dashed lines in Figs.~1,2).
Orbital  excitations are \linebreak 
\\
\begin{table}
\caption{N\'eel order parameter $\langle S^z\rangle $ and some other
expectation values (see text for notations) calculated
in the selfconsistent mean-field approximation ($V=0$),
and corrected by including fluctuation effects ($V\not = 0$).}
\nopagebreak
\begin{tabular}{|c|c|c|c|c|c|c|} %
\tableline 
            &$\langle S^z\rangle$
            &$\langle J_{\perp}\rangle/\langle J_{c}\rangle$
            &$\langle \vec S_{i} \vec S_j \rangle_{c}$
            &$\langle \vec S_{i} \vec S_j \rangle_{\perp}$
            & $E_{mf}$
            & $E_0$\\\hline
$V=0$       & 0.226 
            & 0.052 & -0.417& -0.122 & -0.609   & -0.609\\\hline
$V\not = 0$ & 0.191 
            & 0.072 & -0.421& -0.103 & -0.564   & -0.690 \\\hline
\end{tabular}
\end{table}
\vspace*{-4.0cm}
\begin{figure}
\epsfysize=4.8in
\epsffile{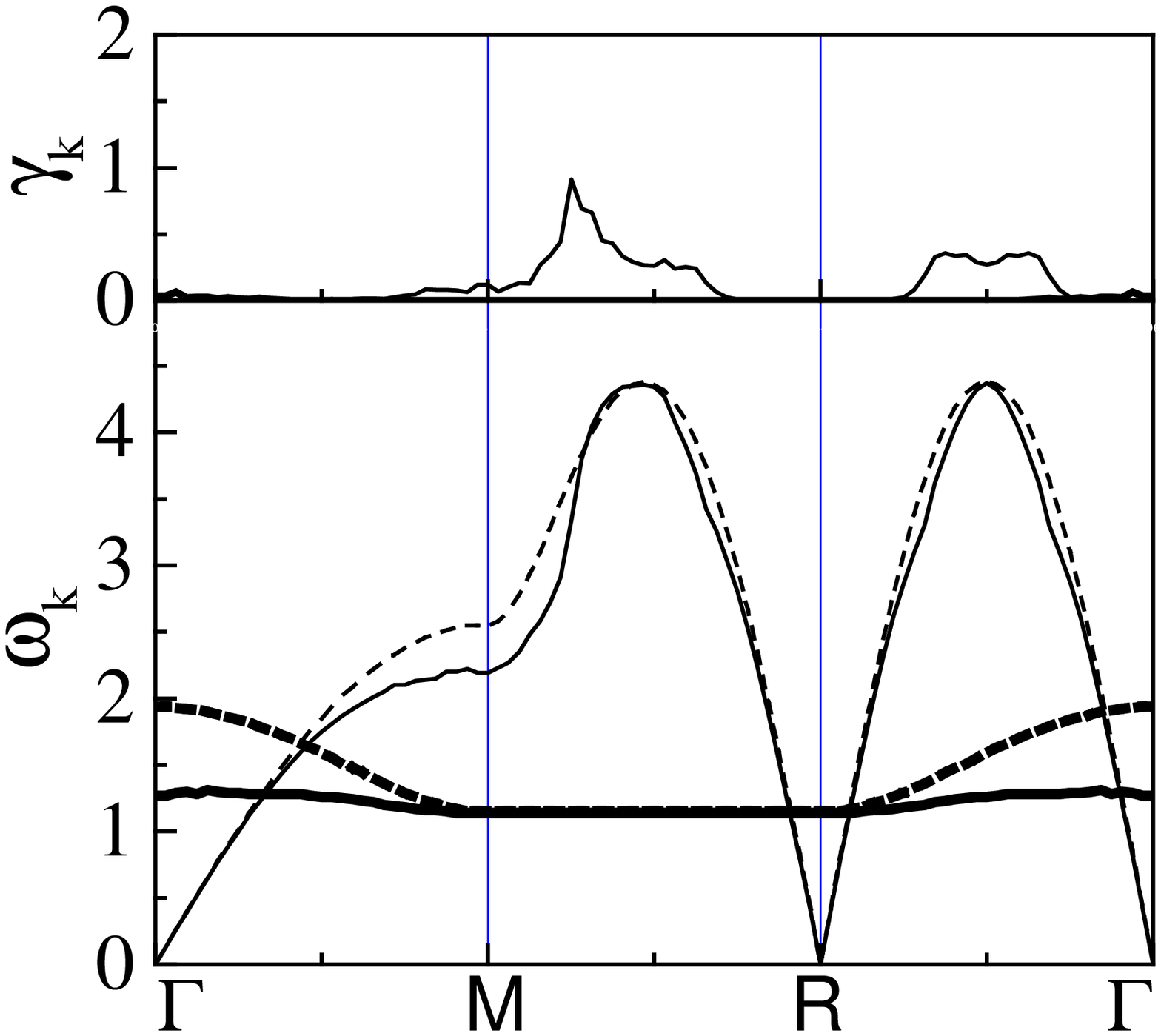}
\vspace*{-1.20cm}
{FIG.~1.  Dispersion $\omega_k $ and damping $\gamma_k$
of the spin and orbital waves along the direction
$\Gamma  \to M(\pi,\pi,0) \to R (\pi,\pi,\pi) \to \Gamma $
 in the Brillouin zone, calculated  including fluctuation effects
(solid lines),  and in the mean-field approximation (dashed lines).
Thin (thick) lines correspond to the spin (orbital) excitation.
$\gamma_k$ for orbital waves is almost indistinguishable
from the zero line.
}\\
\end{figure}
\noindent
gapfull, since the orbital ordering
is not associated with the breaking of any 
continuous symmetry.
Of the similar spirit mean-field picture was recently discussed by
Ishihara et al.~\cite{Ishihara} in context of their
spin-orbital model for manganites.
Quantitatively, we find that the orbital gap is smaller than the
spin-wave bandwidth. The softness of the orbital excitations
is related to that the orbital degeneracy in the model (1) can be
lifted only due to quantum effects in the spin sector. \\

\vspace*{-4.0cm}
\parbox[h]{8.5cm} {
\epsfysize=4.8in
\hspace*{0.0cm}\epsffile{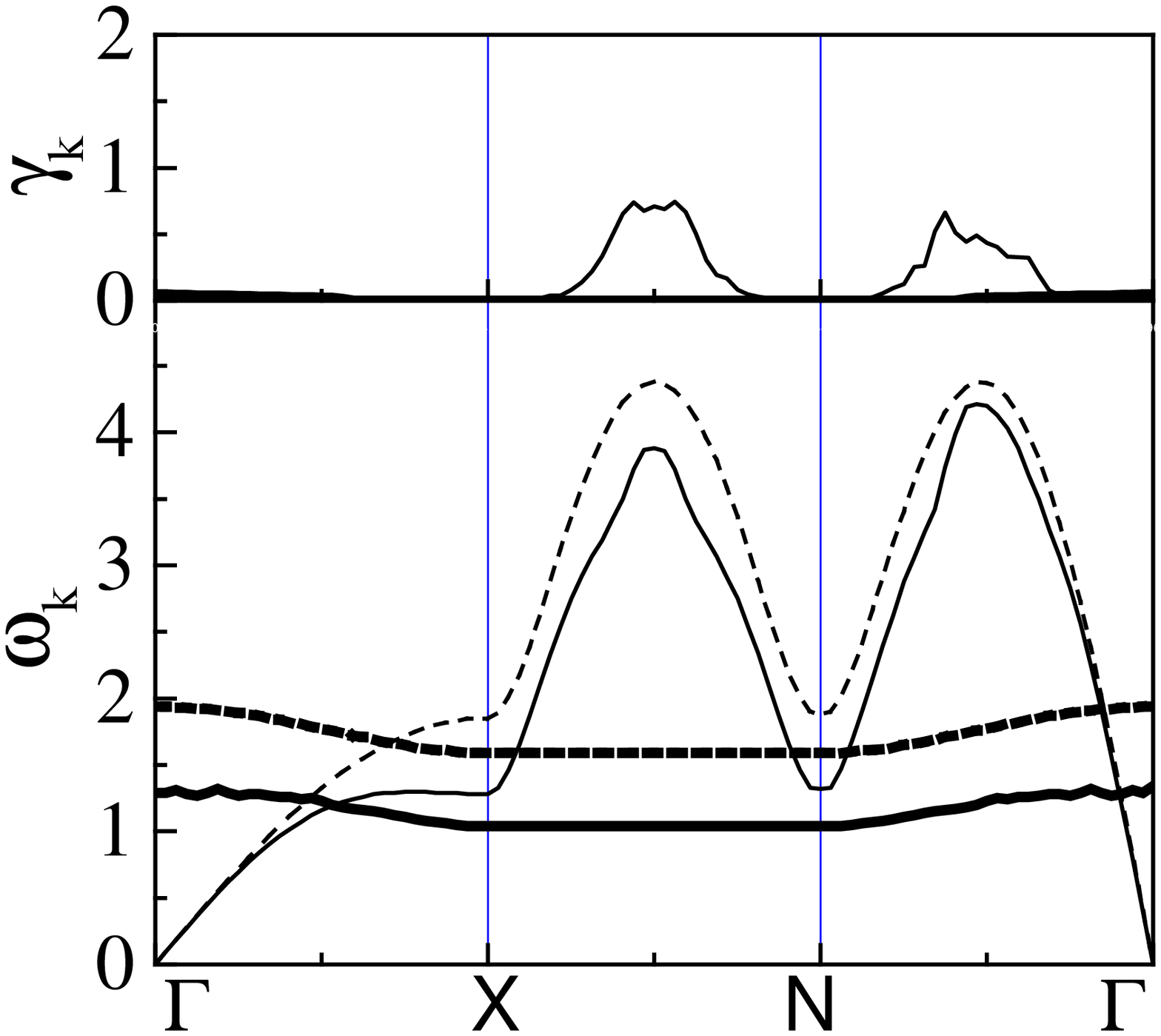}
\vspace*{-1.20cm}
{FIG.~2. The same as  Fig.~1 but along the
$\Gamma  \to X(\pi,0,0) \to N(\pi,0,\pi) \to \Gamma $ direction.  }
}\\

Now, what happens when we switch on  the coupling between spin and
orbital excitations? The latter is represented by
$H_{int}$~(7), which in terms of spin ($\beta_k$) and orbital
($\varphi_k$) wave excitations reads as

\begin{eqnarray}
H_{int}=-V\sum_{kp}^{}&\{ &f_{k,p} \beta_{k}^{+}\beta_{p}+\\\nonumber
& &\frac{1}{2} g_{k,p} (\beta_{k}^{+}\beta_{-p}^{+}+\beta_{-k}^{}\beta_{p})\}
(\varphi_{q} + \varphi_{-q}^{+}),
\end{eqnarray}
where the lowest order (three magnon) terms are only kept.
Here
$ V=\sqrt{3}/2,\;$ and $\; q=k-p$. Matrix elements are
\begin{eqnarray}
f_{k,p}&=&(u_{2q}+v_{2q})[\eta_{q}M_{k,p} +(\eta_{k}+\eta_{p}) N_{k,p}],\\ \nonumber
g_{k,p}&=&(u_{2q}+v_{2q})[\eta_{q}N_{k,p} +(\eta_{k}+\eta_{p})M_{k,p}],\\ \nonumber
M_{k,p}&=&(u_{1k}u_{1p} +v_{1k} v_{1p}), \;\;\;\;
N_{k,p}=(u_{1k}v_{1p} +v_{1k} u_{1p}) ,
\end{eqnarray}
with $\eta_{k}=(\cos k_{x}- \cos k_{y})/2 $.
The Bogoliubov  transformation coefficients in the spin subspace
are given by
$u_{1k}=\{ (s+1)/2 \}^{1/2} $, 
$v_{1k}=-\{ (s-1)/2 \}^{1/2}\mbox{sgn}\gamma_{1k} $,
and
$s=(1-\gamma_{1k}^{2})^{-1/2}$. The factor
$(u_{2k}^{} + v_{2k} ) =(1+2\gamma_{2k})^{-1/4} $
in~(10) is due to Bogoliubov transformation in the orbital sector.

Physically, the interaction (9) accounts for the process when
spin exchange is accompanied by the simultaneous orbital
transition
$\mid z \rangle \leftrightarrow  \mid x \rangle$,
thus  enhancing the $x$ orbital component in the ground state.
Spin-orbital coupling leads to the conventional  $2\times 2$ matrix bosonic
Green's function in both subspaces, with a diagonal ($G$) and
nondiagonal ($F$) components given by\\
\begin{eqnarray}
G_{\omega,k}&=&[(i\omega-A_{\omega,k})+(\omega_k+S_{\omega,k})]/
\mbox{Det}, \\\nonumber
F_{\omega,k}&=&-\Sigma_{\omega,k}^{(a)}/\mbox{Det}, \\\nonumber
\mbox{Det}&=&(i\omega -A_{\omega,k})^{2}- \\\nonumber
&& (\omega_k+S_{\omega,k}-\Sigma_{\omega,k}^{(a)})
(\omega_k+S_{\omega,k}+\Sigma_{\omega,k}^{(a)})
\end{eqnarray}
Here $A_{\omega,k}$ and $S_{\omega,k}$ represent the antisymmetric 
and symmetric (with respect to the Matsubara frequency $i\omega$) 
components of the diagonal self-energy 
$\Sigma_{\omega,k}^{(n)}$ respectively, while 
$\Sigma_{\omega,k}^{(a)}$ is a nondiagonal element of the self-energy matrix.
It is implied that all quantities in Eq.~(11) carry the subspace index
$n$ as well, and $n=1 (2)$ stands for spin (orbital) waves.
We calculate  self-energies from the lowest order diagrams shown in  Fig.~3.
\\  \\

\noindent
\parbox[h]{8.5cm}{
\epsfysize=0.6in
\hspace*{0.0cm}\epsffile{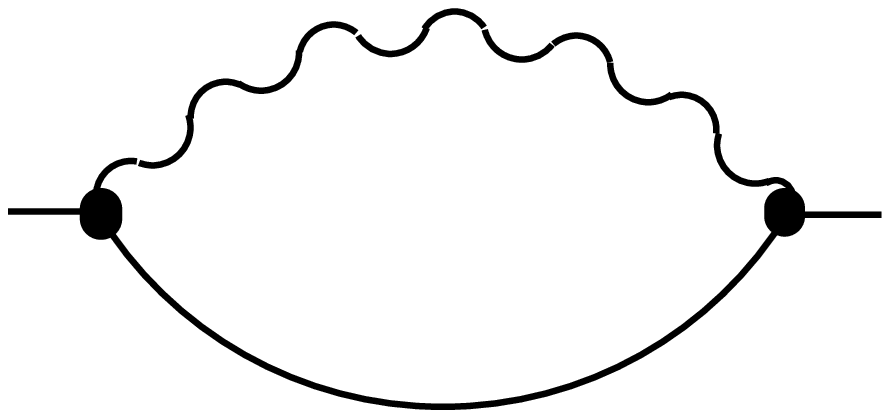}
\epsfysize=0.6in
 \hspace*{0.6cm}\epsffile{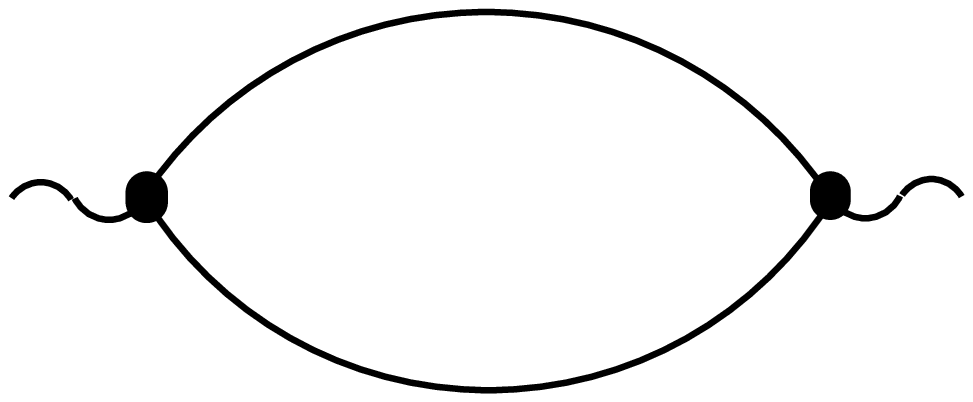}
\epsfysize=0.6in
\hspace*{0.6cm}\epsffile{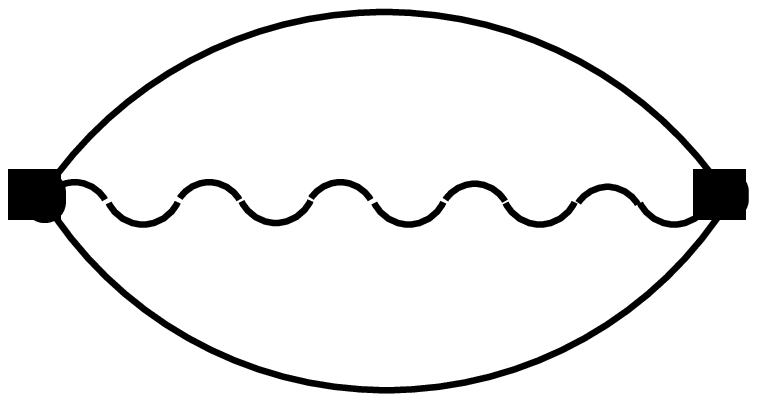}\\[-0.5cm]
{\it \hspace*{1cm} i)\hspace*{2.7cm}ii) \hspace*{2.2cm} iii)}\\
\parbox[h]{8.5cm}{
FIG.~3.
Spin-orbit interaction corrections to  the  spin $(i)$ and
 orbital excitations $(ii)$ , and to the  ground state energy $(iii)$.
 Lines (wavy lines) represent spin  (orbital) waves.
}
}\\[0.4cm]
In spin subspace  we find (at zero temperature):
\begin{eqnarray}
A_{\omega,k}=V^2\sum_{p} (f_{k,p}^{2} - g_{k,p}^{2} )
\dfrac{\omega}{(\omega+i\delta)^{2}-\varepsilon_{k,p}^{2} }\\\nonumber
S_{\omega,k}\pm \Sigma_{\omega,k}^{(a)}
=V^2\sum_{p} (f_{k,p}^{} \pm g_{k,p} )^{2}
\dfrac{\varepsilon_{k,p}}{(\omega+i\delta)^{2}-\varepsilon_{k,p}^{2} }
\end{eqnarray}
Here 
$\varepsilon_{k,p}=(\omega_{1p} +\omega_{2q}),\;\;\; q=k-p $,  and
\begin{eqnarray}
(f^{2}-g^{2} )_{k,p}&=&[\eta_{q}^{2} -(\eta_{k}+\eta_{p})^{2} ]x_{2q},\\\nonumber
(f^{}\pm g^{} )^{2}_{k,p}&=&(\eta_{k} +\eta_{p} \pm \eta_{q})^{2}
(x_{1k}x_{1p})^{\pm 1}x_{2q},
\end{eqnarray}
where 
$x_{1k} =[(1-\gamma_{1k})/(1+\gamma_{1k}) ]^{1/2},\;\;\;
x_{2k}=(1+2\gamma_{2k})^{-1/2}$.

In the orbital sector one finds that $A_{\omega,k} =0$, and
\begin{equation}
S_{\omega,k}=\Sigma_{\omega,k}^{(a)}=2V^{2}\sum_{p}^{} g_{k+p,p}^{2}
\frac{\tilde \varepsilon_{k,p}}
{(\omega+i\delta)^{2}-\tilde\varepsilon_{k,p}^{2} }
\end{equation}
where
$\tilde \varepsilon_{k,p}=(\omega_{1p} +\omega_{1,k+p}) $,
and $g^{2}_{k+p,p}$ can be found from Eq.~(13).
We recall that the ``bare" energies $\omega_{nk}$ in Eqs.~(11)-(14)
are also affected by the interaction, due to the renormalization of parameters
$J_n$ and $\gamma_{nk}$.

Results of self-consistent calculations by including interaction corrections are
presented in Table~1 and Figs.~1,2. Dynamical spin-orbit coupling results in
the following:
$i)$ It enhances quantum fluctuations in both subspaces thus reducing
the staggered moment, and increasing the weight of the $x$ orbital
component (which is about 6 percent). The latter effect is also reflected in
a larger value of the ratio $J_\perp/J_c$ .
$ii)$ Spin and orbital excitations are both softened, which is more 
pronounced in the $k_y=0$ plane (and in equivalent ones), see Fig~2.
The orbital gap still remains well defined. 
A spin-Peierls like instability is absent, 
because of the vanishing matrix elements in Eq.~(9) for momenta along $z$
(note $\eta(0,0,q_z)=0$), and because of the finite interchain coupling. 
$iii)$ Spin waves get a finite damping. Orbital waves are almost undamped
since the density of spin states inside the orbital gap is small.
$i$v) Joint spin-orbital fluctuations significantly lower the ground state
energy (see Table~1).
The latter is given by $E_0=E_{mf}+\langle H_{int}\rangle$,
where the interaction correction to the mean-field result, calculated from
the last diagram in Fig.~3, is
\begin{equation}
\langle H_{int}\rangle=-V^{2}\sum_{k,p}^{}g_{k,p}^{2}/
(\omega_{1k} +\omega_{1p}+\omega_{2,k-p}).
\end{equation}
The exchange energy gain $E_0=-0.69$ per site is found, which is close to our
above estimation from physical considerations.
Summarizing, interaction effects do not qualitatively  change predictions
of the self-consistent mean-field theory, which seems to work quite
reasonably.
This is an important observation giving some credit to the mean-field
ansatz in studying more complicated spin-orbital models.
Of course, the latter fails when the orbital gap is softened close to the phase
boundaries between different orbitally ordered states, and the dynamical 
spin-orbit coupling becomes of  crucial importance.

Considering $x$-type ordering, we found it to be unstable against fluctuations.
It turns out that orbital excitations around this mean-field state are
gapless at the $\Gamma$ point, $\omega_{2q}\sim q$.
In addition, the spin-orbit interaction vertex remains finite at 
$q=0$, since the orbital pseudospin is not conserved quantity, and 
orbital waves can not be considered as Goldstone modes.
All these  lead to the divergencies in perturbation theory indicating that an
$x$-type ordered state is not an appropriate one, as we already mentioned 
above. This result is consistent with ~\cite{Oles}.

In summary, we have studied the spin-orbital coupling problem
in the specific model, where this coupling is particularly important
because of infinite degeneracy of the classical N\'eel state in this model.
The problem of the orbital frustration pointed out in ~\cite{Oles} is actually 
removed by reducing the effective dimensionality of the spin system.
Quantum spin fluctuations then generate an orbital excitation gap through the
spin-orbit coupling mechanism. 
Orbital degeneracy in the model~(1) should manifest itself
in a strong reduction of the N\'eel temperature, by favouring soft
quasi-$1D$ spin structure. 
This is consistent with a basic idea of Feiner at al.~\cite{Oles} that
orbital degeneracy, in general, acts to enhance quantum spin fluctuations.
In contrast to \cite{Oles} we find, however, that this effect is not strong 
enough to destroy the N\'eel order.
Melting of the long-range magnetic order by orbital
fluctuations suggested in \cite{Oles}  does not occur in a cubic perovskite
systems, by simple reason:
A certain (model dependent) orbital ordering always results in the
three dimensional (albeit very anisotropic) network  of exchange
interactions among spins. 
{\em Three dimensionality}  of the spin sector and existence of the 
{\em orbital gap} are important factors stabilizing the N\'eel order.
We believe that the orbital gap is a robust property
of Mott-Hubbard insulators, which is related to the fact that 
the underlying symmetry 
in orbital subspace is only discrete one.
In a metallic state, doped holes can drastically change
the situation, by inducing low-energy orbital fluctuations~\cite{Nagaosa}.
A study of the orbital melting in the Kugel-Khomskii model,
driven by hole doping, deserves further work. 

We would like to thank A.M.~Ole\'s, P.~Horsch and V.~Zevin
for stimulating discussions and useful comments. 
We are also grateful to K.~Davies for a careful reading of the manuscript.
Partial financial support by the Russian State Program
``High-Temperature Superconductivity'', Grant No 95065, and by the Russian
Foundation for Fundamental Research, Grant No 96-02-17527, is acknowledged
by one of us (V.O.)
\hbox{}\\

\end{multicols}

\begin{references}
\bibitem{3} It was found \cite{Oles} that  the Hund's splitting
$J_H$  of the intermediate $d^8$ state  and finite splitting $E_z$
of the $e_g$ orbitals may stabilize  the long range order.
We consider the most ``dangerous" case by setting $J_H=E_z=0$.
\bibitem{Jperp} This approximation is crucial to guarantee that
the interchain correlator 
$\langle \vec S_i \vec S_j  \rangle_\perp \le 0$, and vanishes as
$J_\perp \to 0$.
\end{references}
\end{document}